\documentclass[twocolumn,prb,amsmath,amssymb]{revtex4}
\usepackage{graphicx,bm}

\def\bq{{\mathbf q}}
\def\bj{{\mathbf j}}

\def\bF{{\mathbf F}}

\def\bnab{{\bm \nabla}}

\newcommand{\de}{\delta}

\newcommand{\e}{\epsilon}

\newcommand{\s}{\sigma}
\newcommand{\del}{\nabla}
\newcommand{\p}{\partial}

\newcommand{\bsdel}{\boldsymbol{\nabla}}

\newcommand{\mbj}{\mathbf{j}}

\newcommand{\mbk}{\mathbf{k}}

\newcommand{\mbp}{\mathbf{p}}
\newcommand{\mbps}{{\mathbf{p}\sigma}}

\newcommand{\mbq}{\mathbf{q}}

\newcommand{\mbr}{\mathbf{r}}

\newcommand{\mbv}{\mathbf{v}}

\newcommand{\bsphi}[1]{\boldsymbol{\phi}}
\newcommand{\mbf}[1]{\mathbf{#1}}

\newcommand{\mcal}[1]{\mathcal{#1}}

\newcommand{\bs}[1]{\boldsymbol{#1}}
\newcommand{\nn}{\nonumber\\}
\newcommand{\up}{\uparrow}
\newcommand{\down}{\downarrow}

\newcommand{\bea}{\begin{align}}
\newcommand{\eea}{\end{align}}
\newcommand{\ben}{\begin{equation}}
\newcommand{\een}{\end{equation}}
\begin{document}

\title{Spin Caloritronics  in Noncondensed Bose Gases}
\author{C. H. Wong,* H.J.van Driel, R. Kittinaradorn, H.T.C. Stoof and R.A. Duine}
\affiliation{Institute for Theoretical Physics, Utrecht University, Leuvenlaan 4, 3584 CE Utrecht, The Netherlands}

\begin{abstract}
We consider coupled spin and heat transport in a two-component,  atomic Bose gas in the noncondensed state.  We find that the transport coefficients show a temperature dependence reflecting the bosonic enhancement of scattering, and discuss experimental signatures of the spin-heat coupling in spin accumulation and total dissipation.   Inside the critical region of Bose-Einstein condensation, we find anomalous behavior of the transport coefficients, and in particular, an enhancement for the spin caloritronics figure of merit that determines the thermodynamic efficiency of spin-heat conversion.
\end{abstract}
\maketitle

{The Seebeck and Peltier effects are thermo-electric phenomena that are well understood for ordinary conductors.  Besides being theoretically interesting, these effects have many commercial applications ranging from wine coolers to thermo-electric generators.   Recent developments in spintronics have led to the nascent field of spin caloritronics [\onlinecite{bauerSCC10}] that introduces a spin-dependent generalization of these phenomena.  In solid-state systems, the spin-Seebeck and spin-dependent Seebeck effects have recently been measured [\onlinecite{jaworskiNAT10,slachterNAT10}], while their theoretical explanations are still under active debate [\onlinecite{xiaoPRB10sse}].   

Cold-atom systems  provide a perfectly clean environment to study spin caloritronics, without many of the factors that can make the interpretation of solid-state experiments difficult.  In addition, cold atoms are advantageous in the study of spin and spin-resolved heat transport because one can experimentally realize systems where spin is conserved, apply different temperatures for different atomic species, and measure their distribution functions separately.    While equilibrium properties of these systems have been thoroughly studied, recent research has  started to focus on non-equilibrium behavior such as spin dynamics [\onlinecite{schmaljohannApp04}], heat transport [\onlinecite{meppelinkPRL09}], and spin drag [\onlinecite{vichiPRA99,sommerNAT11}]. The spin drag relaxation rate has been measured in Fermi gases [\onlinecite{sommerNAT11}], and experiments to measure the spin drag conductivity in Bose gases have recently been performed [\onlinecite{koller}]   However, the concomitant thermo-spin phenomena remain largely unexplored.

In this Letter, we study the coupling of spin and heat transport in a cold atomic Bose mixture of two spin species and calculate the spin and heat transport coefficients in the noncondensed state.  In particular, we show that the bosonic nature of the particles leads to qualitatively different temperature dependence of these coefficients as compared to electronic systems.  Furthermore, we introduce a spin caloritronic figure of merit for this system called ``$Z_sT$",  analogous to the ``$ZT$"  figure of merit that determines the efficiency of devices based on the usual solid-state thermo-electric effect.   We computed the temperature dependence of  $Z_sT$  and find  an initial downturn on approach to the critical temperature of Bose-Einstein condensation, followed by an enhancement inside the critical region. This result is interesting, both in the context of developing high efficiency ``atomtronic" devices based on thermo-spin phenomena and the study of dynamical critical phenomena [\onlinecite{halperinPRB75}]   Our theoretical results also have implications for thermo-spin phenomena in other systems where the transport is mediated by degenerate bosons, such as a quasi-equilibrium magnon gas [\onlinecite{demokritovNAT06}].

We begin by considering a cold boson system above the critical temperature  $T_c$ of Bose-Einstein condensation, composed of two different spin {states selected from a larger integer-spin multiplet}, which we will label ``spin up" ($\uparrow$) and ``spin down" ($\downarrow$).     We apply  forces  and temperature gradients which are equal and opposite for the two spin species, i.e., $\bF_\uparrow = -\bF_\downarrow $ and $\bnab T_\uparrow = -\bnab T_\downarrow $ [\onlinecite{relax}].    In response to the ``spin force" and ``spin temperature gradients" defined by $\bF_s\equiv\bF_\uparrow -\bF_\downarrow$ and $\bnab T_s\equiv\bnab T_\uparrow  -\bnab T_\downarrow $, there will be a spin current and spin heat current,  $\bj_{\rm s} = \bj_\uparrow - \bj_\downarrow$, and $\bq_{\rm s} = \bq_\uparrow - \bq_\downarrow$, respectively.  We define the linear-response coefficients by (shown schematically in Fig. \ref{cartoon}):
\begin{figure}[t]
\begin{center}
\includegraphics[width=\linewidth]{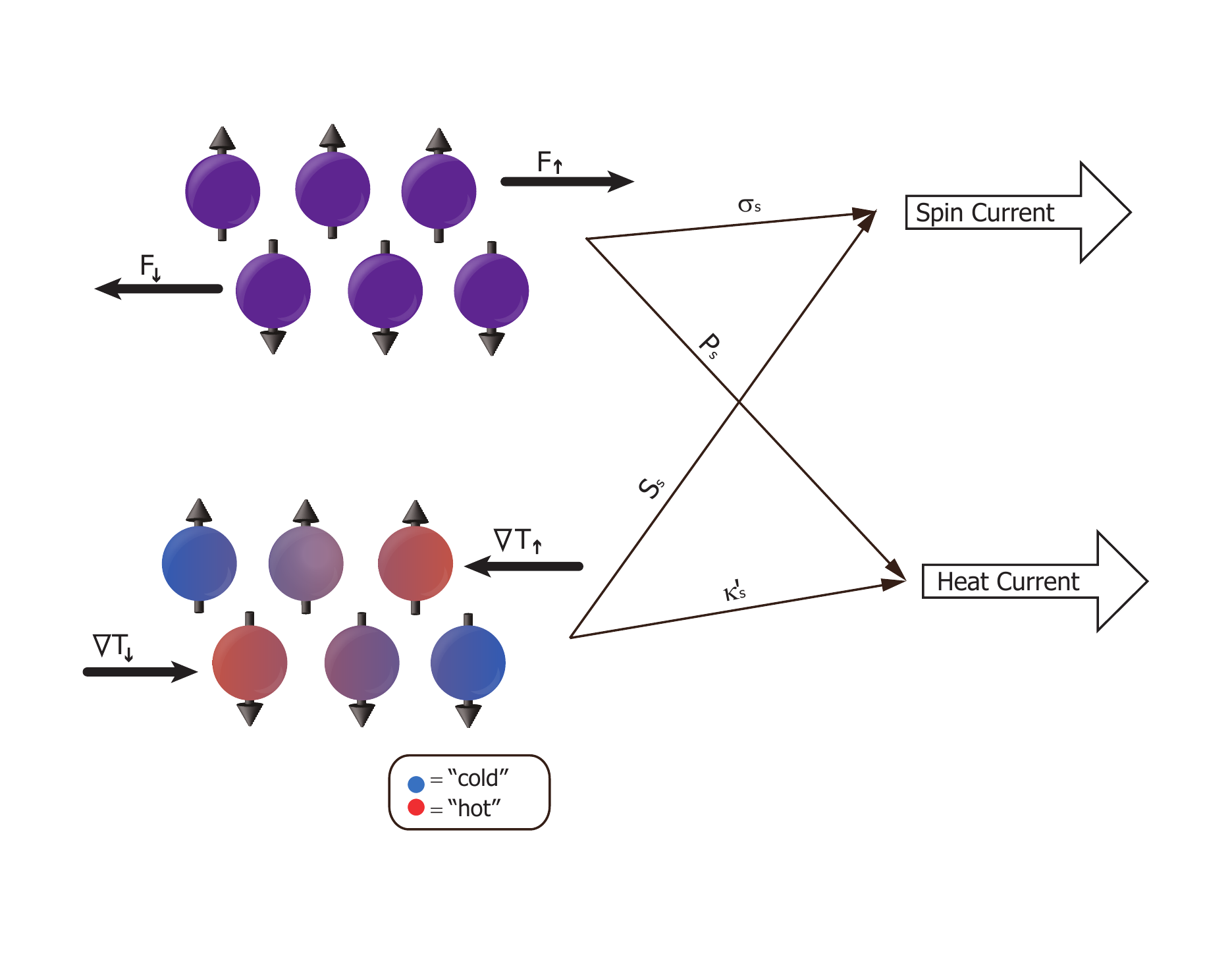}
\caption{{{Schematic illustration of the coupled spin and heat transport.}   Spin-dependent forces or temperature gradients, labeled by $\bF_{\up,\down},\bsdel T_{\up,\down}$, generates both spin and spin heat currents proportional to the coefficients $\s_s$ and $P_s$ in former and $\kappa_s$ and $S_s$ in the latter case, as illustrated in the figure.}
}
\label{cartoon}
\end{center}
\end{figure}

\begin{align}
\left(
  \begin{array}{c}
    \bj_s \\
    \bq_s \\
  \end{array}
\right) \equiv \left(
            \begin{array}{cc}
              \sigma_{\rm s} & \sigma_{\rm s} S_{\rm s} \\
              \sigma_{\rm s} P_{\rm s} & \kappa_{\rm s}' \\
            \end{array}
          \right) \left(
                    \begin{array}{c}
                      \bF_{\rm s} \\
                      -\bnab T_{\rm s} \\
                    \end{array}
                  \right)\,.
\end{align}
In the above, $\sigma_{\rm s}$ is the spin  conductivity, $\kappa_{\rm s}'$ is the spin heat conductivity at zero $\mbf{F}_s$, and $S_{\rm s}$ and $P_{\rm s}$ are the spin-Seebeck and spin-Peltier coefficients, respectively,  which are related by the Onsager reciprocity principle, $P_{\rm s} = S_{\rm s} T$, $T$ being the temperature.  Although we have written the response matrix in a form analogous to the thermo-electric  coefficients defined in metals, we note that here the microscopic mechanisms for the coupling between spin and heat flows are different than those in metals because there is no disorder and lattice phonons in the Bose gas.  The spin-heat coupling studied here is akin to the thermo-diffusion effect in multicomponent classical gases [\onlinecite{smithTP89}].  The spin conductivity is, to the leading order, determined by the viscosity between up and down atoms that arises from inter-spin scattering, which is called spin drag.  In contrast, the spin heat conductivity, which has dependence on intra-spin scattering, is finite even in the absence of inter-spin scattering.

Using the Boltzmann equation for a two-component Bose gas, we have computed $\sigma_{\rm s}$, $\kappa_{\rm s}'$ and $S_{\rm s}$  as a function of $n \Lambda^3$, with $n$ the equilibrium particle density per spin state that is   assumed to be equal for both spin species and $\Lambda = \sqrt{2\pi\hbar^2/mk_BT}$ the thermal deBroglie wavelength, where $m$ is the particle mass, $\hbar$ the Planck's constant, and $k_B$ the Boltzmann constant. The results are shown in Fig.~\ref{Ds} (inset) and \ref{spinheat}.  The order of magnitude of the Seebeck coefficient is $S_s\sim-.01k_B\sim-1 \,\mu\, {\rm eV/K}$, comparable to what is found in ferromagnetic materials [\onlinecite{slachterNAT10}].   Notable is the sharp decrease of all transport coefficients as one approaches the critical value $n\Lambda^3\simeq2.61$.  This effect is due to bosonic enhancement of scattering into occupied states,  which dramatically increases as one approaches the critical temperature $T_c=2\pi\hbar^2/mk_B\Lambda_c^2$ where  $\Lambda_c\simeq(2.61/n)^{1/3}$.  This effect is illustrated in Fig.~\ref{frel}, in which we plot the effective relative momentum distribution $f_{\rm rel}$, defined so that the interspin collision rate is given by ${1}/{\tau_{\rm inter}}=\s_{\rm inter} \int \,p_r^2dp_r v_r f_{\rm rel}(p_r)/ (2\pi\hbar)^3$, where $\s_{\rm inter}=4\pi a^2$ is the interspin scattering cross section and $v_r$ the relative velocity.   As shown in the figure, this distribution increases sharply as one approaches the critical temperature. This is in contrast to spin drag in degenerate Fermi gases, where due to Pauli blocking,  below the Fermi temperature the spin conductivity increases as the temperature is lowered [\onlinecite{bruunPRL08}].
\begin{figure}[t]
\begin{center}
\includegraphics[width=\linewidth]{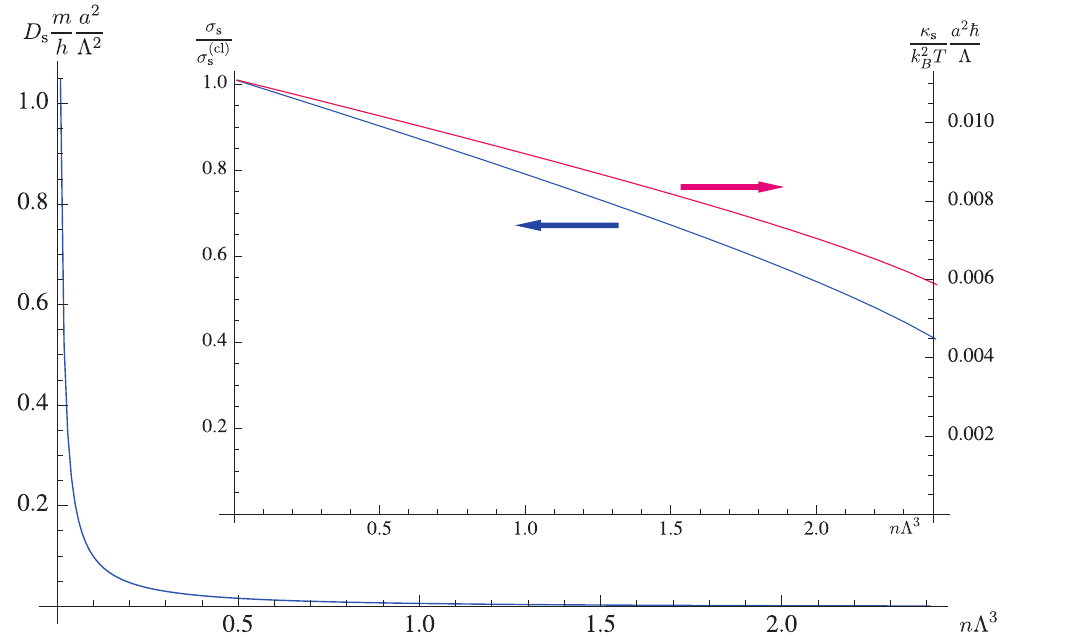}
\caption{Spin diffusivity [normalized by $(h/m)(\Lambda^2/a^2)$] as a function of the degeneracy parameter $n\Lambda^3$.  Inset:  A plot of the spin conductivity $\s_s$ relative to the classical value (left axes)
$\s_s^{\rm (cl)}=(3/64 \sqrt{2} \pi)(\Lambda/ a^2\hbar)$ and a plot of the spin heat conductivity at zero spin current, $\kappa_s$, normalized by $k_B^2T\Lambda/a^2\hbar$ (right axis).}
\label{Ds}
\end{center}
\end{figure}
\begin{figure}[t]
\begin{center}
\includegraphics[width=\linewidth]{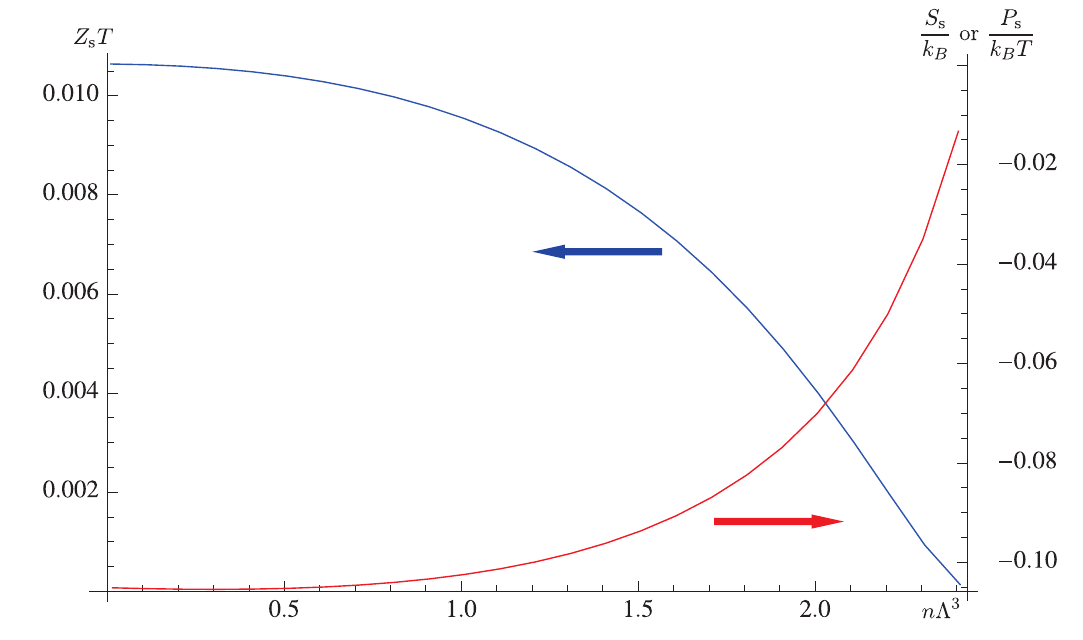}
\caption{ Spin-Seebeck coefficient in units of $k_B$ (right axis) and the  dimensionless figure of merit $Z_sT$ (left axis).  The plots here are outside the critical region, where the absolute magnitude of both coefficients decreases with temperature on the approach to the critical point (Note that the values of $S_s$ are negative).}
\label{spinheat}
\end{center}
\end{figure}

Next we outline the calculation of the transport coefficients outside the critical region.   We start with the two-component, static Boltzmann equation for the semiclassical distribution function $n_\mbps(\mbr)$ under spin-dependent external forces $\mbf{f}_\s$, 
\ben
(\mbv_\mbp\cdot\bsdel+\mbf{f}_\s\cdot\bsdel_\mbp)n_ {\mbps}=C_{\mbps}[n_\up,n_{\down}],
\label{bolt}
\een
where $\mbv_\mbp=\mbp/m$ is the particle velocity, $\s=\up,\down$ the (pseudo)spin of the two-component Bose gas, and the collision integrals $C_{\s}$ are given by 
\begin{align}
&C_{\mbp_1\s}[n_\up,n_{\down}]=-\int{d\mbp_2\over(2\pi\hbar)^3} |\mbv_{\mbp_1}-\mbv_{\mbp_2}|\int  d\Omega'_r \sum_{\tau=\up,\down}{d\s_{\s\tau}\over d\Omega'_r}\nn
&[n_{1 \s}  n_{2 {\tau}}(1+n_{3 \s})(1+n_{4 {\tau}})-n_{3 \s}  n_{4 {\tau}}(1+n_{1 \s})(1+n_{2 \tau})].
\label{coll1}
\end{align} 
The collision integral describes the 2-body elastic scattering of particles labeled by $(\mbp_1\s,\mbp_2\tau)\to(\mbp_3\s,\mbp_4\tau)$, and $\Omega'$ is the solid angle between ingoing and outgoing relative momenta $\mbp_r=(\mbp_1-\mbp_2)/2$ and $\mbp_r'=(\mbp_3-\mbp_4)/2$, respectively.  We take the inter-spin differential cross section to be  $d\s_{\up\down}/d\Omega_{\rm inter}=a^2$, where $a$ is the scattering length, and take the intra-spin terms to be $d\s_{\up\up}/d\Omega=d\s_{\down\down}/d\Omega=2a^2$ on account of Bose statistics [\onlinecite{ferro}].   We  parametrize the non-equilibrium, steady state distribution by 
\ben
 n_ {\mbp\s}(\mbr)=f_ {\mbp\s}(\mbr)-\p_\e f^0_\mbp \phi_{\mbps}(\mbr),
 \een
where $f^0_\mbp=(\exp[(\e_\mbp-\mu)/k_BT]-1)^{-1}$ is the equilibrium Bose distribution, $\mu$ is the chemical potential,  and $f_ {\mbps}(\mbr,t)=(\exp{[\e_\mbp-\mu_\s(\mbr))/k_BT_\s(\mbr)}]-1)^{-1}$  is the local equilibrium distribution, $\p_\e f^0_\mbp=-{f^0_\mbp(1+f^0_\mbp)/k_BT}$, and $\phi_\mbps$ describes the response to the spatial inhomogeneities in $f_\mbps$ and contain all dissipative effects.   This parametrization represents an expansion in the ratio of the mean free path to spatial gradients [\onlinecite{landauPK}].

Linearizing the Boltzmann equations with respect to $\phi_{\mbps}$ and gradients in $f_{\mbps}(\mbr)$, we find that, for our choice of scattering lengths,  the equation for the spin distribution $n_\up-n_\down$ decouples from the equation for the total distribution $n_\up+n_\down$.  
 We consider the linear response of the spin distribution to spin forces $\mbf{f}_s$, gradients in the spin chemical potential $\mu_s$, and spin temperature  $T_s$ independently from the response to the average forces,  i.e., we consider the spin dependent forces  $\mbf{f}_\s= {\s}\mbf{f}_s/2$ and $\bsdel T_\s= {\s}\bsdel T_s/2$.   The linearized drift terms in the left-hand side of Eq.~\eqref{bolt} are proportional to $\bsdel T_s$ and $\bsdel\mu_s$, but they are not independent thermodynamic forces because the chemical potential has dependence on the density and temperature,  $\mu_\s=\mu_\s(n_\s,T_\s)$.    We therefore transform the drift terms using the Gibbs-Duhem relation, $d\mu_s=-sdT_s+(1/n)dp_s$, and the thermodynamic identity $w=\mu+Ts$,  where $s$ is the equilibrium entropy per particle, $w$ is the enthalpy per particle, $p_s=p_\up-p_\down$ is the spin pressure, and we consider the response to the total thermodynamic force $\mbf{F}_s=\mbf{f}_s-\bsdel p_s/n$ and  $\bsdel T_s$.  Writing the linear-response solution as
\begin{figure}[t]
\begin{center}
\includegraphics[width=\linewidth]{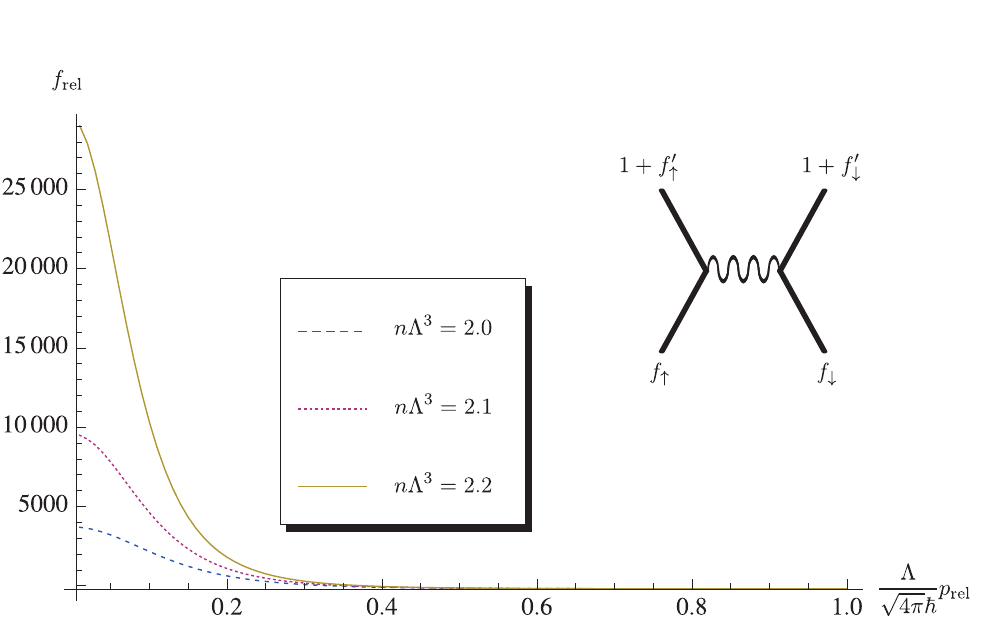}
\caption{A plot of an effective relative momentum occupation, $f_{\rm rel}(p_{\rm rel})$, for three values of the parameter $n\Lambda^3$.   The figure clearly shows a sharp increase in $f_{\rm rel}$ as one approaches the critical temperature.   The Feynman diagram indicates that the interspin scattering rate is increased by the Bose enhancement factors, $(1+f'_\up)(1+f'_\down)$, for the outgoing states.  
}
\label{frel}
\end{center}
\end{figure} 

\ben
\phi_{\mbp s}\equiv\bs{\phi}_{F}(\mbp) \cdot\mbf{F}_s+\bs{\phi}_{T}(\mbp)\cdot(-{\bsdel T_s }),
\een 
the linearized collision integral in the spin equation can be written as $C_{\mbp s}(\phi_s)\equiv C_\mbp(\bs{\phi}_{F}) \cdot\mbf{F}_s+ C_\mbp(\bs{\phi}_{T}) \cdot(-\bsdel T_s)$.  The linearized Boltzmann equation for the spin distribution requires that  $\bs{\phi}_{F}$, $\bs{\phi}_{T}$ satisfy
\ben
\p_\e f^0_\mbp \mbv_\mbp=C_\mbp(\bs{\phi}_{F}) ,\quad
\p_\e f^0_\mbp \left({\e_\mbp-w\over T}\right)\mbv_\mbp=C_\mbp(\bs{\phi}_{T}). 
\label{bolt1}
\een

Using the solution to Eq. \eqref{bolt1}, the spin and spin heat currents are given by
\begin{align}
{\mbj}_s&=-\int{d^3p\over(2\pi\hbar)^3} \p_\e f^0_\mbp\,\mbv_\mbp \phi_{\mbp s}\nn
{\mbq}_s&=-\int{d^3p\over(2\pi\hbar)^3}\p_\e f^0_\mbp  \,(\e_\mbp-w)\mbv_\mbp \phi_{\mbp s}
\end{align}
To solve Eq.~\eqref{bolt1}, we expand the solutions in a power series, $\bs{\phi}_F=\sum_{n=0} a_n(\mu,T)\left({\e_\mbp/ k_BT}\right)^n\mbp$, $\bs{\phi}_T=\sum_{n=0} b_n(\mu,T)\left({\e_\mbp/ k_BT}\right)^n\mbp$, and take moments of the Boltzmann equation by multiplying Eq.~\eqref{bolt1} by $({\e_\mbp/ k_BT})^n\mbp$ and integrating over $\mbp$, resulting in a series of equations of $a_n,b_n$, which we truncate and solve at the second order.  The transport coefficients are readily expressed in terms of the expansion coefficients and the temperature dependence were computed numerically.   The Bose enhancement of the  spin drag rate in the absence of spin heat currents,  calculated in Refs.~[\onlinecite{duinePRL09,DrielPRL10}], was computed using the leading-order solution which describes local Bose distributions for the spin up/down particles rigidly shifted apart,  resulting a spin current.   The second-order solution which we have determined here represents a distortion of the local Bose distribution and is necessary to capture coupled spin and heat flows [\onlinecite{correct}].

We  note that $\s_s$, which determines the spin current driven by external forces, is related to the spin diffusivity $D_s$, which determines the spin current driven by diffusive forces induced by spin density gradients via $\bj_{s}=-D_s\bsdel n_s$, where $n_s=n_\up-n_\down$ is the spin density.  This diffusive current tends to return the system to homogeneous equilibrium and must satisfy the Einstein relation, $D_s=\s_s/\chi_s$, where $\chi_s=\p n_s/\p\mu_s$, is the static spin susceptibility and $\mu_s=\mu_\up-\mu_\down$  is the spin accumulation.  We plot the spin diffusivity in Fig.~\ref{Ds} using the noninteracting spin susceptibility.   The diffusivity also determines the spin density gradient induced by a spin temperature gradient when the spin current is zero.  For example, for a typical density of $n=10^{12}~{\rm cm}^{-3}$ at a temperature of $T=1.5~{\rm\mu K}$ ($n\Lambda^3=2.41$),  we find $|\del n_s/\del T_s|=\chi_sS_s\sim10^{11}~\mu K^{-1} {\rm cm}^{-3}$.  For a typical temperature gradient of $1~{\rm \mu K/mm}$ and for a cloud size of $1{\rm mm}$ [\onlinecite{meppelinkPRL09}], we find a spin density accumulation of $\de n_s=10^{11}~{\rm cm^{-3}}$ , which gives a sizable experimental signal of a $\de n_s/n=10\%$.

Similarly, the heat current driven by $\bsdel T_s$ tends to reduce unequal distributions of energy between spin up and down particles. The total dissipation is given by
 \ben
 T\p_t\mcal{S}={\s_s\over2}\mbf{F}_s^2+{\kappa_s'\over2T}(\bsdel{T}_s)^2+{\s_s S_s}\mbf{F}_s\cdot\bsdel T_s\,,
 \label{diss}
 \een
 where $\mcal{S}$ is the total (spin-summed) entropy.  It is thus possible to measure $\kappa_s'$ and $S_s$ by measuring the heating.  Furthermore, the last term,  analogous  to Thompson heating, is sensitive to the relative signs of $\bF_s$ and $\bsdel T_s$, and allows one to clearly distinguish experimentally the heating contribution from the spin-Seebeck coupling.  It is important to note that typically,  experiments are done in the presence of a trapping potential, which introduces spatial dependence in the transport coefficients.  {The measured values of the transport coefficients should be compared with the trap-averaged values, which differs from the results presented here, but may readily be computed using our Boltzmann formalism.}

For ordinary conductors, one defines a dimensionless figure of merit $ZT = \sigma S^2 T/\kappa$, with $\sigma$ the conductivity, $\kappa$ the heat conductivity in the absence of current and $S$ the thermo-electric Seebeck coefficient, which determines the efficiency of engines based on thermo-electric effects.  Analogously, we define $Z_{\rm s} T = \sigma_{\rm s} S_{\rm s}^2 T /\kappa_{\rm s}$, where $\kappa_{\rm s} = \kappa_{\rm s}' - \sigma_{\rm s} S_{\rm s}^2 T$  is the spin heat conductivity at zero spin current,  which is plotted in the inset of Fig.~\ref{Ds} [\onlinecite{wiede}].   We plot $Z_sT$ as a function of $n \Lambda^3$ in Fig.~\ref{spinheat} , and {observe an initial decrease as one approaches Bose-Einstein condensation.}   The quantity $Z_sT$ measures the ratio between the magnitude of the spin heat coupling and total dissipation, thus the downturn observed results from the faster rate of decrease in the spin heat coupling compared to the total dissipation  as one approaches Bose-Einstein condensation.

However, the results based on the semiclassical Boltzmann equation do not capture the critical fluctuations near the phase transition of Bose-Einstein condensation.   It is known from the theory of dynamical critical phenomena that transport coefficients show anomalous behavior in the critical region [\onlinecite{halperinPRB75}] which is experimentally accessible [\onlinecite{DonnerSCI07}].   In this region, we have computed the transport coefficients using the Kubo formula with the Hamiltonian density $\mcal{H}= \sum_{\s} \psi_\s^\dag\left[-{\hbar^2\over 2m}\bsdel^2-\mu\right]\psi_\s+{1\over2}\sum_{\s\tau}T_{\tau\s}\psi_\s^\dag\psi^\dag_{\tau}\psi_\tau\psi_\s$,
where $\psi_\s$ are the bosonic field operators and the strength of the contact interaction is given by the two-body T-matrix element: $T_{\s\tau}=4\pi\hbar^2a_{\s\tau}/m$.   By neglecting the vertex corrections, the transport coefficients can be expressed in terms of the spectral function $\rho_\s(\mbk,\omega)=-{\rm Im} [G^R_\s(\mbk,\omega)]/\pi\hbar$, which is proportional to the imaginary part of the Fourier transform of the one-particle retarded Greens function, $G^R(\mbr,t;\mbr',t')=i\theta(t-t')\langle[\psi_\s(\mbr,t),\psi^\dag(\mbr',t')]\rangle$.  The spin conductivity then reads [\onlinecite{kittinaradornCM11}]
\ben
\s_s=\frac{\pi\hbar^4}{12m^2k_BT}\int\,{d^3k\over(2\pi)^3}k^2\int\,d\omega \sum_{\s=\up,\down}\frac{\rho^2_\s(\mbk,\omega)}{\sinh^2(\hbar\omega/2k_BT)}
\label{sigma}
\een
while $\s_sS_s$ and $\kappa'_s$ are given by including an additional factor $({\e_\mbp-\mu})/{T}$ and ${({\e_\mbp-\mu})({\e_\mbp-w})}/{T}$, respectively, in the integrand above.
Near the Bose-Einstein phase transition, according to Eq.~\eqref{sigma}, the power law scaling of $\s_s$ follows from the scaling equation for the spectral function, i.e.,
\ben
\rho(\mbk,\omega,\mu-\mu_c)={\lambda^{2-\eta}}\rho(\lambda\mbk,\lambda^z\omega,\lambda^{1/\nu}(\mu-\mu_c))\,,
\een
where $\lambda$ is an arbitrary scaling parameter and $\mu_c$ is the critical chemical potential.  Furthermore, $\eta$ is the anomalous dimension, $\nu$  is the exponent for the correlation length $\xi\sim(\mu-\mu_c)^{-\nu}\sim(T-T_c)^{-\nu}$  and $z$ is the dynamical exponent.  The theory of static and dynamical critical phenomenon [\onlinecite{halperinPRB75}] predicts that the exponents are close to their the mean field values of $\eta=0$, $\nu=1/2$, and $z=2$, in agreement with our numerical calculations.  With these values of the exponents and in three dimensions we find we find an enhancement of the figure of merit which scales as  $Z_{\rm s} T\sim\xi\propto (T - T_c)^{-1/2}$.   {A numerical calculation [\onlinecite{kittinaradornCM11}] shows that the temperature range for this upturn is $|T-T_c|/T_c\approx60(a/\Lambda_c)^2$.  Furthermore, in two dimensions, the critical region given by the Ginzburg criterion is larger, $|T-T_c|/T_c\propto a/\Lambda_c$, while the nature of the phase transition is different, and thus warrants further study.}  Lastly, since this result is generic to systems in the same universality class, we may expect similar behavior in other systems of quantum degenerate bosons, such as quasi-equilibrium magnons.

This work was supported by Stichting voor Fundamenteel Onderzoek der Materie (FOM), the Netherlands Organization for Scientific
Research (NWO), by the European Research Council (ERC) under the Seventh Framework Program (FP7).


\begin{thebibliography}{22}
\bibitem{relax}
{Note that there is no relaxation of the center-of-mass of the atoms because there is no disorder, and in the absence of a confining potential, the system only reaches a steady state with this choice of forces and temperature gradients.}
\bibitem{ferro}
{We note that for this choice of scattering lengths, there is no ferromagnetic phase transition in the normal phase.}
\bibitem{correct}
{The correction to the spin drag previously calculated is  less than $1\%$ and thus negligible.}
\bibitem{wiede}
{ Note that the 
Wiedemann-Franz law that relates the charge conductivity to thermal conductiviy for electrons in solid-state systems is not applicable to our cold-atom Bose system.}
\end{thebibliography}
\end{document}